\documentclass[%
onecolumn,%
oneside,%
floats,%
aps,%
 prd,%
nobibnotes,%
nofootinbib,%
amsmath,%
amssymb,%
amsfonts,%
amscd,%
  superscriptaddress,%
eqsecnum%
]{revtex4}

\usepackage[utf8]{inputenc}
\usepackage{graphicx,array,dcolumn}
\usepackage{cases}
\usepackage[paperwidth=210mm,paperheight=297mm,centering,hmargin=1.8cm,vmargin=2.5cm]{geometry}
\usepackage{enumerate}

\usepackage{hyperref}
\hypersetup{hidelinks}

\usepackage[normalem]{ulem}
\usepackage{soul}

\newcommand{\bi}{\bibitem}

\newcommand{\rar}{\rightarrow}

\def\be{\begin{eqnarray}}
\def\ee{\end{eqnarray}}

\def\-g{\sqrt{-g}}

\renewcommand\rho{\varrho}

\begin{document}

\title{BEASTS IN LAMBDA-CDM ZOO}
\author{\firstname{A.~D.}~\surname{Dolgov}}
\email{dolgov@itep.ru}
\affiliation{Novosibirsk State University, Novosibirsk, Russia}
\affiliation{A.I.~Alikhanov Institute of Theoretical and Experimental Physics, Moscow}
\affiliation{University of Ferrara, Italy
}

\begin{abstract}

Recent astronomical discoveries of supermassive black holes (quasars), gamma-bursters, supernovae, and  dust at high
redshifts, z = (5 --10), are reviewed. Such a dense population of the early universe is at odds with the conventional 
mechanisms of its possible origin. Similar data from the contemporary universe, which are also in conflict with natural 
expectations, are considered too. Two possible mechanisms are suggested, at least one of which can potentially solve all 
these problems. As a by-product of the last  model, an abundant cosmological antimatter may be created.

 \end{abstract}

\maketitle

\section{Introduction \label{s-intro} }

{Observing the sky reveals a great deal of mysteries not yet explained by the known physics.}
{Though the standard cosmological model (SCM) pretty well describes the universe evolution, it does not fit the 
narrow frameworks of the standard model of particle physics and presents clear indications to something new beyond its frameworks.} 
Moreover, even inside the SCM  some internal inconsistencies are constantly arising and they are sufficiently persuasive to demand
serious reconsideration of the basic principles of SCM.

According to the SCM the known world history started from inflation, which has created a suitable for life universe.
In particular, inflation has generated primordial density perturbations at astronomically large scales.
{The calculated spectrum of these perturbations very well agrees with the measured spectrum of angular fluctuations of 
CMB and with the gross features of large scale structure of the universe.} The main ingredients of the universe today are 
the usual baryonic matter, dark matter (DM), and dark energy (DE).
The data allow to conclude that the fractional contributions of different form of matter into the total cosmological energy
density are respectively:
\be 
\Omega_b \approx 0.05, \,\,\,  \Omega_{DM} \approx 0.27, \,\,\, \Omega_{DE} \approx 0.68 .
\label{energy-balance}
\ee 
Dark matter is supposed to be cold (CDM), which means that its free-streaming length is much smaller than the galactic size. It 
consists of nonrelativistic particles whatever they are: new stable weakly interacting massive elementary particles (WIMPs) or 
huge non-luminous astrophysical objects (MACHOs), which can be primordial black holes, low luminosity stars or some 
other astrophysical objects. Dark energy is similar to vacuum energy (or, what is the same, to Lambda term), with negative 
pressure approximately equal by the absolute value to its energy density, $ { P \approx -\rho}$.
That's why the SCM is called ${\Lambda}$CDM cosmology.

{${\Lambda}$CDM   cosmology basically works pretty well} 
{but there are quite a few troubling features, see e.g.~\cite{cdm-probl}:}\\
{1. Cusps in galactic centers, ${ \rho \sim (1/r)^\nu } $ with ${\nu = 1-2}$, which are not observed.  }\\
{2. Too many galactic satellites.}\\
{3. Disk destruction by the CDM clumps.}\\  
4. Much smaller angular momentum of galaxies than observed.\\
{All them can be artifacts of numerical simulations when essential effects from normal dissipating baryonic matter are ignored} 
or be e.g. an indication to existence  of a new form of DM, say, warm DM (WDM).

In addition to these pretty old problems an avalanche of mysteries has emerged 
during the last few years, which can neither be 
helped by WDM, nor by  an improvement of the numerical simulations. It is discovered that 
the standard $\Lambda$CDM cosmology, very well describing
gross properties of the universe, encounters serious problems in numerous
smaller details. The saying goes: "The devil is in the details" but it 
maybe better to say that not the devil but instead "God is in the detail", meaning that these small details 
are celestial indications to  New Physics.

The data came both from the early universe at the redshifts $z\sim 10$ and from the contemporary one. In both cases the objects
have been discovered, which could not be created by the accepted mechanisms based on the known physics.
The young universe at high $z$ and at the age of a several hundred million years
is found to be abundantly populated by quasars, which are presumably the supermassive black holes (BHs)
with the masses up to 10 billions solar masses. Even having in our disposal 13 billion years of the cosmological time,
it is not an easy task to create the observed in the contemporary universe the supermassive  BHs 
(in fact a convincing theory of the creation of supermassive black holes is
absent), but their creation is practically impossible in much younger universe 
assuming the standard mechanism of BH creation by the matter accretion on some preexisting seeds.
In addition to the young quasars
the early universe contains an unexpectedly large number of supernovae, gamma-bursters, very bright galaxies and a lot of
dust. It also looks as "mission impossible" to make all that using the available conventional "instruments". 

Plenty of mysteries are also found in the present day universe.
There are supermassive black holes in 
all large galaxies with not yet established mechanism of their creation. Moreover, such black holes are observed in small
galaxies where they make a too large fraction of the galaxy masses. Such BHs simply cannot be created through the 
mechanism of matter accretion, the only one which is known in the standard case. There is an unexplained accumulation
of quasars in rather small patch in the sky. The origin of MACHOs discovered through microlensing remains mysterious.
The last but not the least, there are several too old stars observed in the Galaxy, at least one of which looking older than the 
Universe. 

Each piece of data may be questioned but taken together they surely show the general tendency in the same direction,
demanding a strong revision of the accepted picture. Almost quoting Marcellus from "Hamlet" we may conclude:
{"Something is rotten in the state of \st{Denmark} the Universe"}.

\section{High $\bf {Z}$ universe \label{s-hi-z}}

\subsection{Universe age as a function of redshift \label{ss-t-of-z}}

There is a large "zoo" of the recently discovered astronomical objects at high-$ z$ which simply could not be created 
in the corresponding surprisingly short times.
To start with, it would be instructive to present the universe age as a function of redshift:
 \be
 t (z) = \frac{1}{H}\,\int_0^{\frac{1}{z+1}} \frac{dx}
{\sqrt{1-\Omega_{tot} +({\Omega_m}/{x}) + x^2\Omega_v } },
\label{t-of-z}
\ee 
where ${\Omega_{tot} = 1}$, ${\Omega_m  = 0.317}$  (it includes densities of baryonic and dark matter),  
and the density of dark energy,  ${\Omega_v = 0.683}$ (sub $v$ here means vacuum-like). $H$ is the present day
value of the Hubble parameter. According to the Planck determination through the angular fluctuations of CMB, 
${ H = 67.3}$ km/sec/Mpc, while the traditional astronomical measurements give a larger value up to  
{${ H = 74}$ km/sec/Mpc. The origin of this discrepancy is unclear~\cite{planck-prmtrs}. If this is not systematics or some
other measurement problem, this tension can be removed if there exists a new unstable DM particle with the life-time
exceeding the universe age at recombination~\cite{BDT-decay}. Depending upon the value of $H$,  the universe age at different $z$ 
is equal to
${ t_U \equiv t(0) = { 13.8/ 12.5}}$;
${ t(12) = { 0.37}/{0.33;}}\,\,$
${ t(10) = { 0.47}/{0.43}}$;
${ t(6.3) = { 0.87}/ {0.79;}}\,\,$
${ t(3) = { 2.14}/ {1.94.}}$
The age is given in Gyga-years and the first higher values of $t(z)$ correspond to smaller value of $H$.

\subsection{Bright galaxies  at high z \label{ss-high-z-gal}}

Several galaxies have been observed at high redshifts thanks to amplification by the natural gravitational lens ``telescopes. 
There is among them  
a galaxy at ${z \approx 9.6}$} which was created when the universe was about 0.5 Gyr old~\cite{gal-9.6}.
Moreover a galaxy at {${z \approx 11}$} has been observed~\cite{gal-11},
which was formed significantly earlier, when the universe age was below $t(11)\approx 0.41$~Gyr, or even shorter, 
$t(11) \approx 0.38$ Gyr, for larger H. Such an early appearance of the galaxies was not expected in the standard
model.

An observation of not so young but extremely luminous galaxy was reported a year ago~\cite{gal-hi-lum}.
Its luminosity and age are estimated as $L= 3\cdot 10^{14} L_\odot $ and  $t  \sim 1.3 $ Gyr respectively. 
It is impressively $10^4$ more luminous than our Milky Way.
{According to the paper, the galactic seeds, or embryonic black holes, around which the galaxy was formed,  
might be bigger than thought possible.}
One of the authors of this work P. Eisenhardt said: "How do you get an elephant?  One way is start with a baby elephant."
But it is unclear how this baby was born. The BH should have been already billions of ${M_\odot}$ , when our
universe was only a tenth of its present age of 13.8 billion years.

Soon after this School was over, another very bright galaxy, GN-$z11$, was discovered at $z=11.1$~\cite{gal-11-2}.
It is remarkably and unexpectedly luminous for a galaxy at such an early time, $t \lesssim 0. 4 $ Gyr. Its UV luminosity
is approximately 3 times larger than that of galaxies observed at $z = 6-8$. There is a large number of 
galaxies, about 800, identified at $z=7-8$. The list of references can be found in the above quoted paper. So the universe was 
rather densely populated when it was younger than $t(6) \approx 0.8  $  Gyr.

As it is asserted in ref.~\cite{monsters}, the presence of GN-$z11$ implies a number density $\sim 10^{-6}\,{\rm Mpc^{-3}}$,
 roughly an order of magnitude higher than the expected value based on extrapolations from lower redshift. According to the
 estimate of this work, based on the existing observations, an  enormous increase in volume which will be surveyed by the new 
 Wide-Field Infrared Survey Telescope  (WFIRST) will provide observations of about 1000 galaxies with $M_{\rm UV} < -22$ 
 beyond $z = 11$ out to $z = 13.5$. But the origin of their creation remains unclear.

According to ref.~\cite{Melia} {"Rapid emergence of high-z galaxies so soon after big bang} 
may actually be in conflict with current understanding of how they came to be." 

\subsection{High redshift quasars \label{|ss-hi-z-QSO}}

Another and even more striking example of early formed objects are high z quasars.
About 40 quasars with ${z> 6}$ are already known, each quasar containing BH with 
${M \sim 10^9 M_\odot}$.  Such black holes, created
{when the Universe was substantially  younger than one billion years,} 
is in strong confrontation with the conventional models of the formation and growth of
black holes and the coevolution of black holes and galaxies. An interesting example is a
quasar which has been observed with maximum {${ z = 7.085}$}~\cite{qso-7085}, i.e. it was 
formed earlier than the universe reached {${t  = 0.77}$ Gyr.} 
Its luminosity and mass are respectively : 
{${L= 6.3 \cdot 10^{13} L_\odot}$} and {${M=2 \cdot 10^9 M_\odot}$,}
{The quasars are supposed to be supermassive black holes}
and their formation in such short time by canonical mechanisms looks problematic to say the least.

Very recently another monster was discovered at  the redshift 6.3 and mass about 12 billions solar masses~\cite{qso-630}.
It has the optical and near-infrared luminosity a few times greater than those of
previously known $z> 6$ quasars.
{There is already the mentioned above  serious problem with formation of lighter and less luminous quasars}
{which is multifold deepened with this new "creature".}
{The new one with ${M \approx 1.2\cdot 10^{10} M_\odot }$  makes the formation}
{absolutely impossible in the standard approach.}

In this connection it is proper to 
continue the quotation from the mentioned in the previous subsection paper~\cite{Melia}:  
"This problem [of  early formed galaxies] 
is very reminiscent of the better known (and probably related) {premature appearance of supermassive
black holes at ${z\sim 6}$.} It is difficult to understand how {${10^9 M_\odot}$} black holes
appeared so quickly after the big bang {without invoking non-standard accretion physics
and the formation of massive seeds,} both of which are not seen in the local Universe." When the quote paper appeared,
the champion quasar with ${M \approx 1.2\cdot 10^{10} M_\odot }$  was not yet discovered.
More discoveries of low luminosity quasars and bright galaxies at $5.7< z < 6.9$
are reported in paper~\cite{qso-gal}.

Some recent papers on possible conventional mechanisms of early black hole formation can be found in ref.~\cite{bh-early}.
At the present stage none of them is particularly persuading. Even the origin of much older supermassive black holes observed 
in the present day universe remains mysterious.

\subsection{Dust, Supernovae, and Gamma-bursters\label{ss-dust}}

The universe at ${z >6}$ is found to be quite dusty, as it is observed in ref.~\cite{dust-1} in numerous galaxies with the 
maximum redshift  $z=6.34$ (galaxy HFLS3), 
and in  ref.~\cite{dust-2} at $z = 7.55$ in the gravitationally-lensed galaxy A1689-zD1, which is
the earliest known galaxy, where the interstellar medium (ISM) has been detected.
This redshift corresponds  to the universe which was only about 500 Myr old.
The results of the last  work are confirmed in ref.~\cite{dust-cnfrm}. 
A catalogue of high-redshift dusty galaxies was created on the basis of observations~\cite{dust-catalogue}, where it was
concluded that the total number of these dusty sources is at least an order of magnitude higher than predicted by
galaxy evolution models. 

The sources of dust in the interstellar medium are stellar explosion.
According to ref.~\cite{dust-source-1}, "even assuming maximally efficient supernova dust production, the observed dust mass of 
the z = 7.5 galaxy A1689-zD1 requires very efficient grain growth. This, in turn, implies that in this galaxy the average density 
of the cold and dense gas, where grain growth occurs, is comparable to that inferred from observations of QSO host galaxies 
at similar redshifts. Although plausible, the upper limits on the dust continuum emission of galaxies at 
$6.5 < z < 7.5$ show that these conditions must not apply to the bulk of the high redshift galaxy population."

In reference~\cite{dust-source-2}, in addition to supernovae, another possible dust source was studied, namely   
asymptotic giant branch (AGB) stars. The conclusion was also pessimistic that the AGB stars are not numerous and efficient enough
to create the observed amount of dust at $z = 4 - 7.5$. 
"Supernovae could account for most of the dust, but only if all of
them had efficiencies close to the maximal theoretically allowed value. This
suggests that a different mechanism is responsible for dust production at high
redshifts, and the most likely possibility is the grain growth in the
interstellar medium."

Probably the amount of supernovae in the early galaxies is considerably larger than expected.
Another argument in favor of this conclusion it that the medium around the observed early quasars contains
considerable amount of ``metals'' (elements heavier than He). 
According to the standard picture, only elements up to ${^4}$He  { and traces of Li, Be, B}
were formed by BBN, {while heavier elements were created
by stellar nucleosynthesis and} {dispersed in the interstellar space by supernova explosions.}
Hence, prior to or simultaneously with the QSO formation a rapid star formation should take place.
{These stars should evolve to a large number of
supernovae enriching interstellar space by metals through their explosions} which later make molecules and dust.
(We all are dust from a neighboring supernova explosion. but probably at much later time.) 
Another possibility to enrich the universe with metals  
is a non-standard BBN in bubbles with  very high baryonic density, which allows for formation of heavy elements 
beyond lithium~\cite{bbn-hi-B}, see below.

Observations of gamma ray bursters (GBR) also  indicate at
{an immense  abundance of supernova at large redshifts.} 
{The highest redshift of the observed GBR is 9.4} and there are a few more
GBRs with smaller but still large redshifts. 
{The necessary star formation rate for explanation of these early
GBRs is at odds with the canonical star formation theory.} 
But despite an imperfection  of the  theory, such high-z objects do exist, and thus we have to conclude
that prior to or simultaneously with the QSO formation a rapid star formation should take place.
These stars could produce plenty of supernovae which enriched interstellar space   by metals.
A possible mechanism of GBR generation at high $z$ through such early star collapse 
is considered in ref.~\cite{GRB-gen}. 

In the model, which is discussed in this lecture below, such stars and heavy primordial black holes could be  abundantly 
enough created in the very early universe.

\section{Contemporary or near-contemporary universe \label{s-contemp-univ}}

\subsection{ Old stars in the Milky Way \label{ss-old-stars}} 

With an increased precision of the nuclear chronology the ages of several stars have 
been recently determined to be much older than expected. Below we quote a few recent results.

Employing thorium and uranium  abundances
in comparison with each other and with several stable elements {the age of
metal-poor, halo star BD+17$^o$ 3248 was estimated as
${13.8\pm 4}$ Gyr~\cite{cowan}.
For comparison the age of the inner halo of the Galaxy is } {${11.4\pm 0.7}$ Gyr~\cite{kalral}.

The age of a star in the galactic halo, HE 1523-0901, was estimated to be 
about 13.2 Gyr~\cite{frebe}.
First time many different chronometers, such as the U/Th, U/Ir, Th/Eu and Th/Os ratios to
measure the star age have been employed.

The metal deficient {\it high velocity} subgiant in the solar neighborhood
HD 140283  has the age ${14.46 \pm 0.31 }$ Gyr~\cite{H-Bond}.
The central value exceeds the universe age by two standard deviations,
if ${H= 67.3}$ and ${t_U =13.8}$, while for  ${H= 74}$, and thus ${ t_U = 12.5}$ the excess is by nine
standard deviations. So this star really looks older than the universe.

A possible explanation of this discrepancy could be an unusual initial chemical content of such stars.
Normally a pre-stellar cloud consists of 25\% of $He^4$ and 75\% of hydrogen. However, in the scenario discussed below
the primordial nucleosynthesis is able to create much heavier elements, as is mentioned in the previous sections
or/and the early supernovae could enrich the interstellar gas with heavy elements, so the initial chemical 
content of some stars would be much different from the traditional one. Such stars could evolve to their
present state considerably faster than the usual ones. Of course this conjecture should be verified by
the stellar nucleosynthesis calculations.

In addition to these very old stars a planet in the Milky Way was discovered~\cite{old-planet}. 
with the age  $10.6^{+1.5}_{-1.3} $ Gyr. For comparison the age of the Earth is "only" 4.54 Gyr.
To create such a rocky planet SN explosion must precede  its formation and this takes also a
non-negligible time.

\subsection{ Supermassive black holes today \label{ss-sprm-BH}}

{The astronomical observations very strongly suggest  that every large galaxy and some smaller 
ones contain a central supermassive BH} with masses 
larger than { ${ 10^{9}M_\odot}$} in giant elliptical and compact lenticular galaxies
and {${\sim10^6 M_\odot}$} in spiral galaxies like Milky Way.
The mass of the central BH is typically about 0.1\% of the mass of the stellar bulge~\cite{BH-bulge},
but some galaxies may  have huge  black holes: e.g. NGC 1277  has
the central BH  of  ${1.7 \times 10^{10} M_\odot}$, or ${60}$\% of its bulge mass~\cite{NGC1277}.
The origin of these  superheavy BHs is not understood.
These observational data create serious problems for the
standard scenario of formation of central supermassive BHs by accretion of matter in the central 
part of a galaxy. An inverted picture looks more plausible, when first a supermassive black hole was formed and 
attracted matter serving as a seed for subsequent galaxy formation~\cite{AD-JS}. 

More  examples of the same kind can be found in ref.~\cite{khan-2015}. As the authors say,
although supermassive black holes  correlate well with their host
galaxies, there is an emerging view that outliers exist.
 Henize 2-10, NGC 4889,
and NGC1277 are examples of super massive black holes  at least an order of magnitude more massive
than their host galaxy suggests. 
{The dynamical effects of such ultra-massive central black holes is unclear. }

{A  discovery of a very compact dwarf galaxy
older than 10 Gyr, enriched with metals, and probably with a massive black in its center} 
also seems to be at odds with the standard model~\cite{strader}.
The dynamical mass of the galaxy is ${2\times 10^8 M_\odot}$ and the radius is
${R \sim 24}$ pc, so its density is very  high.
The Chandra data reveal a variable central X-ray source with $L_X \sim 10^{38}$ erg/s that could be an active galactic nucleus 
associated with a massive black hole or a low-mass X-ray binary. Analysis of optical spectroscopy shows that the object is quite  
old, $ \gtrsim 10$ Gyr, and has more or less the solar metallicity, with elevated [Mg/Fe] and strongly enhanced [N/Fe] that indicates light element 
self-enrichment; such self-enrichment may be generically present in dense stellar systems. 

In the recent paper~\cite{small-gal-BH} an over-sized black hole was observed in a modest mass galaxy. The black hole 
mass in the AGN of the galaxy, was  deduced to be equal to ${M_{BH} = (3.5 \pm 0.8) \cdot 10^8 M_\odot}$ with the 
accretion luminosity ${ L_{AGN} = (5.3 \pm 0.4) \cdot 10^{45} }$erg/s ${\approx 10^{12} L_\odot}$, which is equal to
12\% of the Eddington luminosity. All that is much more than expected for a galaxy of such modest size. The data are in 
tension with the accepted picture in which
this galaxy would recently have transformed from a star-forming disc galaxy into an
early-type, passively evolving galaxy.

Recently an observation of a quasar quartet embedded in giant nebula was reported~\cite{quartet}
in a survey for Lyman-emission Lyman-emission at redshift  ${z \approx 2}$. According to the authors,
it reveals rare massive structure in distant universe. As it is stated in the paper,
all galaxies presumably once passed through a hyperluminous quasar phase powered by accretion onto a 
supermassive black hole. But because these episodes are brief, 
{quasars are rare objects separated by cosmological distances, so
the chance of finding a quadruple quasar is ${\sim 10^{-7}}$.} 
It implies that the most massive structures in the distant universe have a tremendous supply 
(${\sim 10^{11} M_\odot}$) of cool dense (${ n \approx 1/}$cm$\bm{^3}$) gas,
{in conflict with current cosmological simulations.}

All these observations much better fit the inverted scenario mentioned at the beginning of this section.

\subsection{Near-solar mass black holes and MACHOs \label{ss-machos}}

The mass distribution of black holes observed in the Milky way seems to demonstrate some peculiar features not understood by
the conventional theory. It is found that their masses are concentrated in the narrow range
${ (7.8 \pm 1.2) M_\odot }$~\cite{M-BH-narrow}.
This result agrees with another paper where
a peak around ${8M_\odot}$, a paucity of sources with masses below
 ${5M_\odot}$, and a sharp drop-off above
${10M_\odot}$ are observed~\cite{kreidberg}. Such facts are indeed very strange, if these BHs were formed
by the stellar collapse, as it is usually assumed.
 {Astronomical data also hints to  a two-peak mass distribution of the PBHs and compact stars,} which is 
probably observed, but not explained up to now~\cite{farr}. Quoting this work:
"sample of black hole masses provides strong evidence of a gap between the maximum neutron star mass and the lower bound on 
black hole masses". These results may fit the mass distribution of our model of Sec.~\ref{s-explain} 
assuming that 
the lower mass BH are also created by a normal mechanism of stellar collapse.

A similar or maybe even connected problem is related to the nature of MACHOs discovered through 
gravitational microlensing by Macho and Eros groups.
They are invisible (very weakly luminous or even non-luminous) objects with
masses about a half of the solar mass in the Galactic halo and in the center of the Galaxy and recently in
the Andromeda (M31) galaxy. Their density is  significantly greater than the density expected from the known low luminosity
stars but not enough to explain all dark matter (DM) in the halo. The present day situation with MACHOs is briefly reviewed
in ref.~\cite{BDK}, which we follow below.

MACHO group \cite{MACHO2000,Bennet2005}  has announced 13 - 17 microlensing events in the Large Magellanic Cloud (LMC),
a much larger number than expected if MACOSs would be normal weakly shining stars.
The fractional contribution of the density of  these compact "lenses" with respect to the total halo density (which is essentially
the dark matter density) was estimated  to be in the range $ 0.08<f<0.50 $ (95\% CL) for 
$ 0.15M_\odot < M < 0.9M_\odot $. 

EROS (Exp{\'e}rience pour la Recherche d'Objets Sombres) collaboration  has placed only an upper
limit on the halo fraction, $f<0.2$ (95\% CL) for { the} objects in the specified above MACHO
mass interval, while EROS-2 \cite{EROS2007} gives $ f<0.1$ in the mass range $10^{-6}M_\odot<M<1M_\odot$.

AGAPE collaboration \cite{AGAPE2008}, working on microlensing in M31 (Andromeda) galaxy, finds
the halo { Macho} fraction in the range $0.2<f<0.9$,
while MEGA group marginally conflicts with them with an upper limit $f<0.3$
\cite{MEGA2007}.

Detailed analysis of the controversial situation with the results of different groups is given in
ref.~\cite{Moniez2010}.
Newer results \cite{OGLE2013} for EROS-2 and OGLE
(Optical Gravitational Lensing Experiment)
in the direction of the Small Magellanic Cloud are:
 $ {f <0.1} $  at 95\% confidence level for { Macho}s with the mass $ 10^{-2} M_\odot$
and $ {f <0.2} $ for { Macho}s with the mass $ 0.5 M_\odot $.

Thus MACHOs for sure exist. Their density is comparable to the density of the halo dark matter but their 
nature is unknown. They could be brown dwarfs,  dead stars, or primordial black holes. The first two options are
in conflict with the accepted theory of stellar evolution, if MACHOs were created in the conventional way.

\section{Possible explanations \label{s-explain}}

The  described above problems in the sky both in the early, $z\sim 5-10$, and the present day universe
strongly suggest that there exist some new effects outside the standard approach to the theory of formation
and evolution of celestial bodies, so the latter 
demands an essential modification. One possibility is an unusual cosmological expansion 
law~\cite{DHT,melia-t},  such that at these high redshifts the universe happened to be older than in the standard regime.

Another option is an efficient creation 
of stellar-like objects (compact stars,  ancestors of supernovae, primordial black holes, 
including the supermassive ones)  in the very early universe after the QCD phase transition. The necessary for that large density
perturbations on cosmologically small but astrophysically significant scales 
can originate in a well known scenario of the supersymmetric, or  Affleck-Dine~\cite{Affleck}, 
mechanism of baryogenesis after a simple modification. This modification was  
suggested long ago~\cite{AD-JS} and now permits to explain all the described above anomalies in a unique way.

Shortly the scenario is the following.  In supersymmetric  Grand Unified models there exists a scalar  field, $\chi$, with non-zero
baryonic number, {${ B\neq 0}$.} The potential of $\chi$ generically has some flat directions. In a toy model the potential can be 
presented as:
\be
U_\lambda(\chi) = \lambda |\chi|^4 \left( 1- \cos 4\theta \right).
\label{U-of-chi}
\ee
The $\chi$-bosons may condense along flat directions, $\cos (4\theta) = 1$, of this quartic potential. There also exists the quadratic
potential which may appear as a result of some symmetry breaking and is essential at small $\chi$. It can be taken in the form:
\be
U_m( \chi ) = | m_1 |^2 |\chi |^2   - \frac{1}{2} \left( m_2^2 \chi^2 + m_2^{*2} \chi^{*2} \right) =
| m |^2 |\chi|^2} {\left[{ 1-\cos (2\theta+2\alpha)} 
\right],
\label{U-m-of-chi}
\ee
where ${ \chi = |\chi| \exp (i\theta)}$ and ${ m=|m|e^\alpha}$.
{If ${\alpha \neq 0}$, both C and CP are  explicitly broken.}  We took $|m_1| = |m_2|$, though it is not necessary.

{In GUT SUSY baryonic number is naturally non-conserved}. It is described here by non-invariance of ${U(\chi)}$
with respect to global $U(1)$  phase rotation, $\chi \rar \chi \exp (i\alpha)$. 

Initially (after inflation) ${\chi}$ was naturally away from the origin and when 
inflation is over it starts to evolve down to the equilibrium point, $\bm{\chi =0}$,
according to equations of the Newtonian mechanics:
\be
\ddot \chi +3H\dot \chi +U' (\chi) = 0.
\label{ddot-chi}
\ee
with the Hubble friction term $ 3 H\dot \chi$.

Baryonic number  of $\chi$, $B_\chi =\dot\theta |\chi|^2 $, in this language 
is analogous to mechanical angular momentum. 
The rotational degree of freedom in the course of ${{\chi}}$  decay transferred
its baryonic number to that of quarks in B-conserving process. Since quite large baryonic number could be accumulated in 
such rotation, the Affleck-Dine  baryogenesis might create the cosmological baryon asymmetry of order of unity, much larger
than the canonical value ${\sim 10^{-9}}$. 

If ${ U_m (\chi) \neq 0}$, the angular momentum, B, would be generated by a mismatch of the directions
of the  quartic and quadratic valleys, when $\chi$ evolved down to zero from some large values.
{If CP-odd phase ${\alpha}$ is non-vanishing, then both baryonic and 
antibaryonic regions are possible with dominance of one of them.}
{Matter and antimatter domain may exist but globally ${ B\neq 0}$.}
 
 {A minor modification of AD-scenario can lead to very early formation of compact stellar-type objects and
naturally, though not necessarily, to a comparable amount of anti-objects,}
{such that the bulk of baryons and possibly antibaryons (in equal or comparable amount)
would be in the form of compact stellar-like objects or primordial black holes (PBH),} 
{plus sub-dominant observed homogeneous baryonic background.}
{The amount of baryonic matter and antimatter in such compact objects may be comparable or even larger 
than the amount of the { observed}  baryons,} but such compact (anti)baryonic objects
would not contradict  any existing observations.
To this end one needs to add to the usually accepted potential of the
Affleck-Dine field $ \chi$ a general renormalizable coupling to  inflaton $ \Phi$~\cite{AD-JS}; it is the last term 
in the equation below:
\be 
U = \lambda |\chi|^4 \,\ln( \frac{|\chi|^2 }{\sigma^2} 
+\lambda_1 \left(\chi^4 + h.c.\right) + (m^2 \chi^2 + h.c.). 
+ {g|\chi|^2 (\Phi -\Phi_1)^2} .
\label{U-of-chi-Phi}
\ee
If $\Phi $ is close to $\Phi_1 $, then the window to the flat directions, is open but only during 
a relatively short period. However, it could be sufficiently long, so that the bubbles with large
values of $\chi$ would be created and hence huge baryon asymmetry even up to  $\beta \sim 1$ would arise inside them.
Because of the inflationary expansion after $\Phi$ passed through $\Phi_1$
the bubbles could easily become astrophysically large, but still the bubbles would occupy a minor fraction of the 
total volume of the universe. Despite that the mass density of these B-bubbles with high $\beta$  can make a dominant
contribution to the total cosmological matter balance of baryons due to their huge baryon density which may be by far larger than
the observed tiny value of the cosmological baryon asymmetry, ${{ \beta \approx 6\cdot 10^{-10}}}$.  

The distributions of B-bubbles over their radius and mass have log-normal form:
\be
\frac{dN}{dM} = \mu^2 \exp{[-\gamma \ln^2 (M/M_0)]},
\label{dn-dM}
\ee
where ${\mu}$, ${\gamma}$, and ${M_0}$ are some constant parameters, $M$ and $\mu$ have dimensions of mass, while
$\gamma$ is dimensionless. The spectrum is practically model independent, because it is essentially determined by inflation. 

Immediately after formation of the bubbles with large value of ${\chi}$ some inhomogeneities in the energy density were developed 
due to different equations of state inside and outside of the bubbles. The matter was less relativistic inside the bubbles than
outside. But these inhomogeneities remain relatively small. There are large isocurvature perturbations due to much larger
baryon asymmetry inside the bubbles. These isocurvature perturbations are transformed into large density perturbations,
but at small scales, after the QCD phase transition, which took place at $ T= (200 - 100)$ MeV, after massless quarks combined forming
non-relativistic protons. Depending upon the value of the Jeans mass of B-bubbles they could make either primordial
black holes (PBH) or dense compact stellar-like objects. The masses of them are expected to be in the range form a fraction of the
solar mass up to millions solar masses and even higher.
Due to  the subsequent matter accretion, the log-normal mass distribution would be distorted amplifying the high mass tail of the
distribution. The PBHs created through such mechanism can naturally explain the mentioned in subsection~\ref{ss-machos}
features of the several-stellar-mass black holes in the Galaxy.

A modifications of interaction potential between $\Phi$ and $\chi$ leads to a more intricate mass spectrum of the early formed stellar
type objects, e.g.,  if:
\be
U_{\rm int}  = {\lambda_1 |\chi|^2 } 
 \left( \Phi - \Phi_1\right)^2   \left( \Phi - \Phi_2\right)^2 ,
\label{U-int-2}
\ee
we come to a two-peak mass distribution of the PBHs and compact stars, which is 
probably observed, but not explained up to now, if the BHs are dominantly created by the stellar collapse.

\section{Conclusion \label{s-conclud}}

According to the suggested mechanism compact stellar-like objects and primordial black holes could be created in the 
very early universe when it was only ${ t= 10^{-5} - 10^2}$ second old. Their masses should be of the order of the mass of the
matter inside the cosmological horizon, $m_{hor} \approx 10^5 M_{\odot} (t/$sec). As is well known, if the mass created by the
density contrast is equal to the mass inside  the horizon a primordial black hole would be created~\cite{zeld-nov-BH}. For smaller masses 
stellar type objects would be  formed.

The mass distribution (\ref{dn-dM})
is rather strongly cut at high mass tail but it can be enhanced by  matter accretion on lower mass PBHs. In such a way
one can easily populated the universe at $z=5-10$ by the observed objects,  such as high-redshift SN, gamma-bursters,
including 10 billion solar mass quasars, which are very difficult, if possible at all, to create  in other way. The big bang nucleosynthesis
inside or in vicinity of the high-B bubbles creates heavy elements by far more efficiently that the standard BBN. This may lead to 
the observed evolved chemistry and a lot of dust at $z\sim 10$. The unusual primordial chemical content may explain the
existence of stars which are formally older than the universe.

The B-bubbles which did not became PBHs would form very early stars. Such ``stars'' might evolve quickly and, in
particular, make early SNs, enrich the universe with heavy
(anti)nuclei and later reionize the universe.

The early formed superheavy BHs can be the seeds for the galaxy formation. This explains in particular, an existence of the observed superheavy
BHs in small galaxies.  Some or majority of these early stars may be dead by now and are observed as MACHOs.

Since the sizes of these objects are small at the cosmological scale they are neither forbidden by the data on the 
angular fluctuations of CMB nor by its frequency spectrum. The energy release from stellar like objects in the early
universe is small compared to the CMBR energy density.

The anomalous abundances of light elements created by BBN at high $\beta$ are not dangerous for 
the observed abundances  since the anomalous part of the universe volume is small. It is tempting to prescribe 
the Lithium problem if the  metal-poor halo stars, where the anomalously low abundance of Li is observed, were
created from the primordial clouds with abnormal $\beta$.

A natural, though not obligatory, outcome of this model is a prediction of cosmic antimatter, mostly in the
form of compact high velocity stars in the Galaxy. The observational bounds on the abundance of such stars
are quite vague, so we may have a lot of antimatter practically at hand. Such early formed stars would behave 
as cold dark matter and so should have much larger velocity than the normal stars and hence they would populate the 
galactic halo, see ref.~\cite{anti-BB}.

However, it is unclear, if the CDM-problems mentioned at the beginning can be solved, provided galaxies
are formed around earlier created seeds. 

\vspace{0.5cm}
\noindent
{\bf Acknowledgement}\\
This work was supported by the grant of the President of Russian Federation for the State support of the
leading scientific Schools of Russian Federation, number NSh-9022-2016.2.

\end{document}